\documentclass[manuscript]{aastex}
\usepackage{color}

\shorttitle{two jets}
\shortauthors{Chen et al.}

\begin{document}


\title{A Complex Solar Coronal Jet with Two Phases}


\author{Jie Chen\altaffilmark{1,2} Jiangtao Su\altaffilmark{1,3} Yuanyong Deng\altaffilmark{1}  and E. R. Priest\altaffilmark{4}}
\altaffiltext{1}{Key Laboratory of Solar Activity, National Astronomical Observatories, Chinese Academy of Sciences, Beijing 100012, China}\email{chenjie@bao.ac.cn}
\altaffiltext{2}{State Key Laboratory for Space Weather, Center for Space Science and Applied
Research, Chinese Academy of Sciences, Beijing 100190, China}
\altaffiltext{3}{School of Astronomy and Space Sciences, University of Chinese Academy of Sciences, 19A Yuquan Road, Beijing 100049, China}
\altaffiltext{4}{Matematical Institute, University of St Andrews, North Haugh, St andrews, KY16 9SS, UK}



\begin{abstract}
Jets often occur repeatedly from almost the same location. In this paper, a complex solar jet was observed with two phases to the west of NOAA AR 11513 on July 2nd, 2012. If it had been observed at only moderate resolution, the two phases and their points of origin would have been regarded as identical. However, at high resolution 
we find the two phases merge into one another and the accompanying footpoint brightenings occur at different locations. The phases originate from different magnetic patches rather than being one phase originating from the same patch. Photospheric line of sight (LOS) magnetograms show that the bases of the two phases lie in two different patches of magnetic flux which decrease in size during the occurrence of the two phases. Based on these observations, we suggest the driving mechanism of the two successive phases is magnetic cancellation of two separate magnetic fragments with an opposite polarity fragment between them.

\end{abstract}





\section{Introduction}
Jets are transient impulsive phenomena with plasma motions along particular directions (Shibata et al. 1992, 1994; Canfield et al. 1996;  Moore et al. 2010; Sterling et al. 2015). They are ubiquitous in the solar atmosphere and occur in active regions, the quiet Sun, and polar regions 
(Cirtain et al. 2007; Zhang et al. 2014, 2016; Shen et al. 2012; Liu et al. 2014;  Hong et al. 2016; Raouafi et al. 2016). The typical length of jets is 10 - 400 Mm, the width is 5 -100 Mm, and the apparent velocity is 10 - 1000 km s$^{-1}$ with an average value of 200 km s$^{-1}$ (Shimojo et al. 1996, 1998).  

It is generally believed that solar jets are caused by magnetic reconnection. However, the jet characteristics largely depend on the magnetic environment at the jet source. Heyvaerts et al. (1977), Forbes and Priest (1984), Shibata et al. (1992), Moore et al. (2013) presented a standard model for the formation of a jet. The essential idea of this model is that at the base of a jet during the emergence of a magnetic arch into an ambient unipolar field, a current sheet forms between the ambient field and one leg of the arch with opposite polarity. During the emergence, the current sheet becomes larger and thinner, until magnetic reconnection is initiated. Interchange reconnection between closed field and open field occurrs at the current sheet. The hot plasma on the open field lines escapes out to become the spire of the jet and the hot plasma on the closed field lines is seen as a bright point at the jet base. This model has been developed further using 2D numerical 
experiments by Yokoyama and Shibata (1995, 1996), Nishizuka et al. (2008), Jiang et al. (2012), and more recently 3D experiments by Moreno-Insertis and Galsgaard (2013),  
N\'{o}brega-Siverio et al (2016), Archontis and Hood (2013), Syntelis et al (2015), Raouafi et al (2016).

Jets are associated with both flux emergence and cancellation (Chae et al. 1999; Brooks et al. 2007). After new flux has emerged, one or other flux element often cancels with a nearby one of opposite polarity. Most of them occur in mixed polarity regions (Shimojo et al. 1996).  Jets are commonly associated with magnetic cancellation (Muglach 2008, Young  and Muglach 2014a, b, Liu et al. 2011). Jets  tend to occur repeatedly (Chifor at al. 2008; Chen et al. 2015, Liu, et al. 2016) with a period ranging from tens of minutes to hours. It is normally thought that repeated jets occur at the same location with almost the same configuration (Guo et al. 2013; Schmieder et al. 2013). 

In this paper, for the first time, we show that at high resolution a jet can have two different phases involving reconnection of different magnetic regions and we propose 
a model for such a jet.

\section{Instruments}
The observations used in this paper are taken by the Atmospheric Imaging Assembly (AIA; Lemen et al. 2012) and 
Helioseismic and Magnetic Imager (HMI, Scherrer et al. 2012) on board the SDO satellite. Full-disk EUV images supplied
by the AIA instrument have a cadence of 12 s and a spatial resolution of $0.6^{\prime\prime}$. Full solar disk line of sight (LOS) magnetograms 
in the photosphere obtained by HMI have a cadence of 45 s and a spatial resolution of  $0.5^{\prime\prime}$. The aia\_prep.pro is used to process 
AIA level 1.5 data and HMI LOS data. After the reduction of this routine, the data of AIA and HMI will have the same spatial resolution and the full-disk images 
will be co-aligned. It is not safe to reach conclusions about alignment to within $1^{\prime\prime}$ - $2^{\prime\prime}$ purely from the default SSW software alone because of thermal effects that can 
cause changes in alignment between two different telescopes (Orange et al. 2014). However, in the present study we align HMI and AIA 1600 images 
over a broader field of view (about $100^{\prime\prime} \times 100^{\prime\prime}$), in order to include the region around the sunspot. This plage does indeed align well with the 
default SSW alignment, and so we are confident about our AIA 171 alignment.

\section{Observations and Results}


A series of jets was observed on the west edge of AR 11513 on 2012 July 2 (Chen et al. 2015). 
A jet with two phases occurring around 22:30 UT is studied. The peak of Phase 1 was around 22:28 UT. Phase 2 followed Phase 1 quickly and the peak was 
around 22:32 UT. The configurations of Phase 1 and Phase 2 are shown in Figures 1a and 1d with the Phase 2-spire being almost parallel to Phase-1 spire. However, the images demonstrate clearly that the roots of Phase 1 and Phase 2 are located in different places. The root of the Phase 1-spire originates from the southwest of the jet-base, which is marked by R1, while the root of Phase 2-spire originates from the northeast of the jet-base, which is marked by R2. The distance between R1 and R2 is around 2.8  Mm. The two spires also appear in different locations: the spire of Phase 2 is to the northeast of Phase 1-spire about 2.8 Mm.

The evolution process from Phase 1 to Phase 2 is shown in the animation. Originally, there is a bright patch at the jet-base which is located between the negative magnetic fragment `n0' and the positive fragment `p1' (Figs. 1b and 1g). Then the spire of Phase 1 becomes longer and longer, and also has a lateral motion to the northeast. Around 22:28:11 UT, we can see clearly the jet-spire (Fig. 1c) and the bright patch becomes smaller and weaker. Around 22:29:23 UT, the root of the jet varies from southwest 
to northeast. From this time, Phase 1 changes to Phase 2 and during Phase 2 the jet continually moves to the northeast laterally. At 22:31:23 UT, there is a bright patch between `n0' and `p2' to the northeast of the jet-base (Figs. 1e and 1g). At 22:32:11 UT, the whole spire of Phase 2 can be seen with its root located in the northeast (Figs. 1d and 1f).

At the jet-base, there are mixed magnetic polarity regions. In the middle, there is a negative magnetic polarity region which is marked by `n0' in Fig. 1g. To the right of `n0', there is a positive magnetic polarity region marked by `p1'; to the left of `n0', there is another positive magnetic polarity region 
marked by `p2'. n0 and p1 had existed for a long time, but during the occurrence of the jet p1 and n1 decreased, while p2 and n2 flowed out from the penumbra 
and then moved toward n0. The evolution of the LOS magnetograms is shown in the animation. The magnetic fragments n0, p1 and p2 are moving magnetic features. Below the jet the magnetic field is continually changing. 
From HMI LOS magnetograms in Figs. 1g, 1h and 1i, we can see the way the regions n0, p1 and p2 are reducing. Combining images from AIA and HMI, we find that Phase 1 originates from a region between the negative polarity region n0 and the positive polarity region p1 while Phase 2 originates from the region between n0 and p2. 

 In Figure 1, the jet and the bright patches of the footpoints of the two phases are appear in AIA 171 \AA. Actually, the jet and the bright patches are seen in six wavelengths: 171 \AA, 193 \AA, 211 \AA, 335 \AA, 94 \AA, 131 \AA. The configurations of the two phases in AIA 131 \AA\ are shown in Figure 2; their configurations are similar to that in 
 171 \AA. The normalized intensity variations of the bright patch (red box in the image 131 \AA\, Fig. 2a) in 
Phase 1 in six wavelengths are shown in Figure 2a. From 22:24 UT to 22:30 UT, there are three obvious peaks in the six wavelengths. This implies that  the bright patch becomes brighter three times during the growth of Phase 1, and the bright patch has a wide temperature range as shown in the six wavelengths of AIA. The normalized intensity variation of the Phase 2 bright patch (red box in the image 131 \AA, Fig. 2b) in six wavelengths is shown in Figure 2b. From 22:30 UT to 22:34 UT, the trends of intensity variation are same, and there is an obvious peak in the different wavelengths. The peak time of this bright patch is the same as Phase 2 of the jet.

In order to know whether there is magnetic cancellation or not, the variation of magnetic flux is calculated. Figure 3 shows the variation of the magnetic 
flux in the rectangular regions including p1 and p2. From Fig. 3a, we can see that the magnetic flux of p1 decreases from 22:15 UT to 22:24 UT. At 22:26 UT, there is an obvious brightening at the base of Jet 1. Magnetic cancellation between regions n0 and p1 appears to drive Phase 1. The net increase of magnetic flux of p1 from 22:24 UT to 22:45 UT despite the continued cancellation is due to two small positive polarity patches (highlighted with two green circles in Fig. 3b) moving into the p1 region. The variation of magnetic flux in region p2 is shown in Fig. 2b: it decreases from 22:15 to 22:30. At 22:31 UT, there is a brightening at the base of the jet, 
suggesting that flux cancellation between the regions n0 and p2 drives Phase 2.

The motions and magnetic evolution of p1 and p2 are shown in Figure 4. We take a slit between n0 and p1 (Fig. 4a) and plot a time-distance image (Fig. 4c).
Figure 4c shows there is magnetic cancellation between p1 and n0. The magnetic flux of p1 decreases from 22:00 UT to 23:00 UT. Another slit between p2 and n0 
is also taken which is marked by a green line in Figure 4b. The time-distance image is given in Figure 4d, and shows that p2 moves 
towards n0 from 22:00 UT to 22:30 UT.




\section{Interpretation and Discussion}

The formation and evolution of two successive phases of a jet have been observed here at high spatial resolution. 
The two phases overlap in time, with a time between their peaks of only 4 minutes. 
The distance between the peaks of Phase 1 and Phase 2 is only 2.8 Mm. The two patches 
p1 and p2 are very close, if the spatial resolution were low, they would not easily be separated from each other.  At 
high resolution we find Phases 1 and 2 have different configurations: the root of Phase 1 
is in the southwest, the root of Phase 2 is in the northeast. The locations of brightening of the feet of two phases are 
different: the location of the brightening of the footpoint of Phase 1 is between p1 and n0; the location of brightening 
of the footpoint of Phase 2 is between p2 and n0. The two phases and the brightening of their footpoints appear over 
a wide temperature range. We suggest that the two phases are driven by magnetic  
flux cancellation but they originate from different nearby
locations rather than originating from the same place.

The jets occurred to the west of NOAA AR 11513 in a moat region surrounding a sunspot of negative polarity and containing mixed polarity 
moving magnetic features. Thus the ambient magnetic field is predominantly open and directed away from the sunspot. Within this field, small 
positive magnetic fragments such as p1 and p2 tend to form closed flux that is connected to nearby negative fragments. As they move around and 
come closer to other negative fragments, cancellation and reconnection naturally occur to form new connections, jets and heating. Thus, when 
different positive fragments cancel with the same negative flux they tend to produce jets with successive phases that have similar paths due to the nature of the 
ambient field. There is also a big coronal hole to the west edge of 
NOAA AR 11513 with open negative magnetic field (Figure 1 of Chen et al. 2015).

Based on our observation and analysis, we interpret the physical mechanism for the two successive phases 
with a two-dimensional sketch (Figure 5). Figure 5a 
depicts the magnetic configuration before the jet. It is dominated by an open magnetic field 
with negative polarity, which is rooted partly in the region n0 in our observation and partly in n1 and n2. To the right of n0, there is 
a closed magnetic field, with its positive footpoint p1 linked to part of another nearby negative fragment. 
On the left side, there is another closed magnetic structure symmetrically placed with its positive footpoint p2 linking to part of another 
nearby negative fragment. During 
the first process of magnetic cancellation (Figures 5b and 5c), the footpoint motion of p1 and n0 drives magnetic reconnection. Such magnetic 
reconnection between open and closed magnetic field is called interchange 
magnetic reconnection (Moore at al. 2013). The result of the reconnection is to create  
new closed magnetic flux between n0 and p1, and new open magnetic flux in the faraway 
negative polarity region n1. The hot plasma trapped in the new closed 
field region creates a bright patch between n0 and p1, and hot plasma escapes out along the new open field 
to become the spire of Phase 1. The spire is collimated by the dominant strong open magnetic flux. Heat from the reconnection is 
magnetically confined to a small volume in the closed field below the reconnection location and so shows up as a brightening, but in 
the open field above the reconnection site it is spread over a much larger volume and so does not lead to a brightening in region n1. The continuous magnetic cancellation makes the jet move sideways.  It can be seen, by comparing Fig. 5b and 5c, that the location of the open field line that has just reconnected 
(and therefore the spire) moves to the left as reconnection proceeds. This transverse motion can be faster than the photospheric cancellation speed, since 
the transverse motion of photospheric sources can lead to a build up of a current sheet which subsequently reconnects. Phase 2 is driven by the subsequent cancellation between n0 and p2 (Figure 5d), and produces a spire and a brightening in 
the same manner as in Phase 1. It is possible that the two phases are independent of one another, but this is unlikely since they occur so close in time and space and the spire has a more-or-less smooth motion, which suggests that the first phase helped trigger the second phase or that the fragments p1 and p2 are not 
independent but are linked somehow by a common convection pattern in the moat below the photosphere.

The formation and evolution process of the jet is a process in which magnetic energy is converted to kinetic energy and thermal energy via magnetic reconnection. 
From our observations, we estimate the magnetic energy, kinetic energy and thermal energy. For Phase 1 during magnetic cancellation from 22:15 UT to 22:24 UT, the magnetic flux decreases by 1.5$( \pm0.2)\times10^{18}$ Mx in about 9 minutes (Fig. 3a), resulting in a mean magnetic cancellation rate of 
$(2.8\pm0.3)\times10^{15}$ Mx s$^{-1}$. The length of region p1 is around $6.0\times10^8 $ cm.
We calculate the mean magnetic energy release rate during the time of magnetic cancellation to be $(2.8\pm0.1)\times10^{27}$ erg s$^{-1}$, which is similar to the change of magnetic energy in a reconnection event described by Li et al. (2016). Following the calculation of the plasma density and 
temperature by Liu et al. (2016), the average electron number density is about $10^9$ cm$^{-3}$ and the mean temperature is 1.5 MK for the jet.  For the bright patch at the footpoint of the jet, the density is about $2.5\times10^9$ cm$^{-3}$ and the temperature is 4 MK. For the jet, the velocity is about 200 km s$^{-1}$, the width is about 4 Mm and the length is about 30 Mm, and we estimate the mass of jet to be $3.7\times10^{11}$ g. The change of kinetic energy is $7.5\times10^{25}$ erg $s^{-1}$. Likewise, we estimate the change of thermal energy of the jet spire, to be  $7.8\times10^{25}$ erg $s^{-1}$. For the bright patch at the footpoint of 
the jet, the length is about  $8\times10^{8}$ cm and the thermal energy  $7.1\times10^{26}$ erg $s^{-1}$. About 30\% of the magnetic energy is converted to kinetic and thermal energy, so the magnetic energy is sufficient to provide the thermal energy and kinetic energy
of the jet (more details of the calculation can be found in the Appendix).



\acknowledgments

We are thankful for discussion with Dr. Qingmin Zhang. 
SDO is a mission for NASA's Living With a Star program.
This work was partly supported by National Natural Science Foundation of China (grant
Nos. 11303048, 11673033, 11373040, 11427901). This work was also partly supported by an International Exchanges cost share award with 
NSFC for overseas travel between collaborators in the UK and China, and State Key Laboratory for Space Weather, Center for Space 
Science and Applied Research, Chinese
Academy of Sciences.


\appendix

\section{Appendix - Energy Estimation}
Here, we describe the estimates of magnetic energy, kinetic energy and thermal energy for Phase 1 of the jet.

The magnetic flux $\Phi$ decreases  by 1.5$(\pm0.2)\times10^{18}$ Mx in about 9 minutes from 22:15 UT to 22:24 UT (Fig. 3a). 
The magnetic flux density B can be expressed as
\begin{equation}
B=\frac{\Phi}{S},
\end{equation}
where $\Phi$ is the variation of magnetic flux and S the area of magnetic cancellation. The length of the region p1 (L) is around $6.0\times10^8$ cm, and we assume the width is one percent of the length (namely, $6.0\times10^6$ cm). The variation of magnetic flux density B is around 420 G during the period of magnetic cancellation.

The mean magnetic energy release rate $E_m$ during the time of magnetic cancellation (t) can be written as
\begin{equation}
E_m=\frac{B^2}{8\pi\times t}V,
\end{equation}
The length of the region p1 is around $6.0\times10^8 $ cm (L), so the volume V is  L$^{3}$ and $E_m$ is $(2.8\pm0.1)\times10^{27}$ erg s$^{-1}$. 

The change of kinetic energy is
\begin{equation} 
E_k=\frac{1}{2} m v^{2}, 
\end{equation}
where m is the mass of the accelerated plasma in Phase 1. Here m is set to be $\mu m_{0} n_{e} \pi L(d/2)^2 $, where $\mu$ is the mean molecular weight 0.58 for fully ionized coronal plasma, $m_{0}$ is the mass of protons, the width d is about $4\times10^8$ cm, the length L is about $3\times10^9$ cm, and
$n_{e}$ is the estimated average electron number density of Phase 1. The plasma density $n_{e}$ for Phase 1 is about $10^9$ cm$^{-3}$, and the velocity $v$ is about 200 km $s^{-1}$. Using these parameters, we obtain the the mass of accelerated plasma in Phase 1 as $3.7\times10^{11}$ g and the change of kinetic energy as $7.5\times10^{25}$ erg $s^{-1}$. 

The thermal energy of the jet is 
\begin{equation} 
E_{th}=N\kappa_{p}T, 
\end{equation} 
where $\kappa_{p}$ is the Boltzmann constant, N = $n_{e}$V is the number of electrons involved in Phase 1, and V is the volume. T is the average temperature,
which we take to be 1.5 Mk. The thermal energy of the jet spire is $7.8\times10^{25}$ erg $s^{-1}$. For the bright patch at the footpoint of 
the jet, the density is about $2.5\times10^9$ cm$^{-3}$, the temperature is 4 Mk,  and the length is about  $8\times10^{8}$ cm, so the thermal energy is 
$7.1\times10^{26}$ erg.

The mean magnetic energy release rate is $2.8\times10^{27}$ erg s$^{-1}$, while the sum of the kinetic energy and thermal energy of the jet spire is $1.5\times10^{26}$ erg, and the 
thermal energy of the bright patch at the footpoint is $7.1\times10^{26}$ erg. Thus, about 30\% of the magnetic energy is converted to kinetic and thermal energy.

Note that for the Phase 1 of the jet, the average electron number density $10^9$ cm$^{-3}$ and the mean temperature 1.5 MK.  For the bright patch at the footpoint of the jet, the density  $2.5\times10^9$ cm$^{-3}$, and temperature 4 MK are taken from Liu et al. (2016).

\clearpage





\begin{figure}
\epsscale{0.85}
\plotone{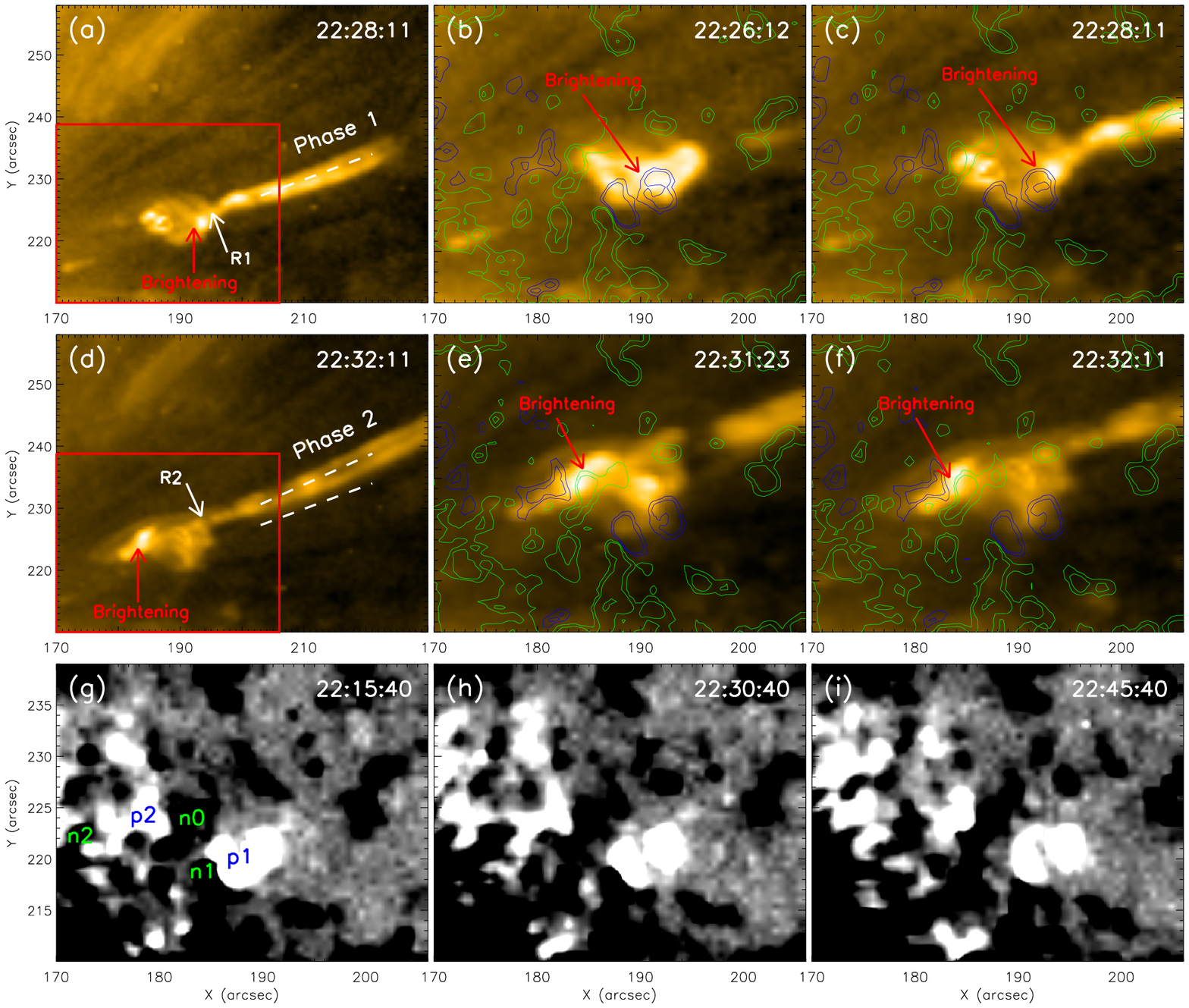}
\caption{(a) An image of Phase 1 of the jet obtained from SDO/AIA 171 \AA\ at 22:28:11 UT on July 2nd, 2012. The red box is the field of view (FOV) for (b) and (c).  (b)-(c) Close-up SDO/AIA 171 \AA\ images near the base of Phase 1 at 22:26:12 UT and 22:28:11 UT overlaid with simultaneous line-of-sight (LOS) magnetic field. Blue/green lines mark contour levels for the positive/negative polarities. (d)  An image of Phase 2 obtained from SDO/AIA 171 \AA\ at 22:32:11 UT. The red box marks the FOV of (e) and (f). (e)-(f) Close-up images near the base of the jet in Phase 2 at 22:31:23 UT and 22:32:11 UT overlaid with simultaneous LOS magnetic field.  (g)-(i) 
SDO/HMI LOS magnetograms at 22:15 UT, 22:30 UT, 22:45 UT,  separately. Dynamic range for the magnetic field strength is 30 G. The white/black color represents positive/negative polarity.   
The FOV is the same as the FOV in (b), (c), (e) and (f). Negative polarity region in the middle of FOV marked by n0,  positive polarity regions to the both sides of n0 marked by p1 and p2, separately. n1 and n2 are also marked in Fig. 1g. ) \label{fig1}}
\end{figure}

\begin{figure}
\epsscale{0.85}
\plotone{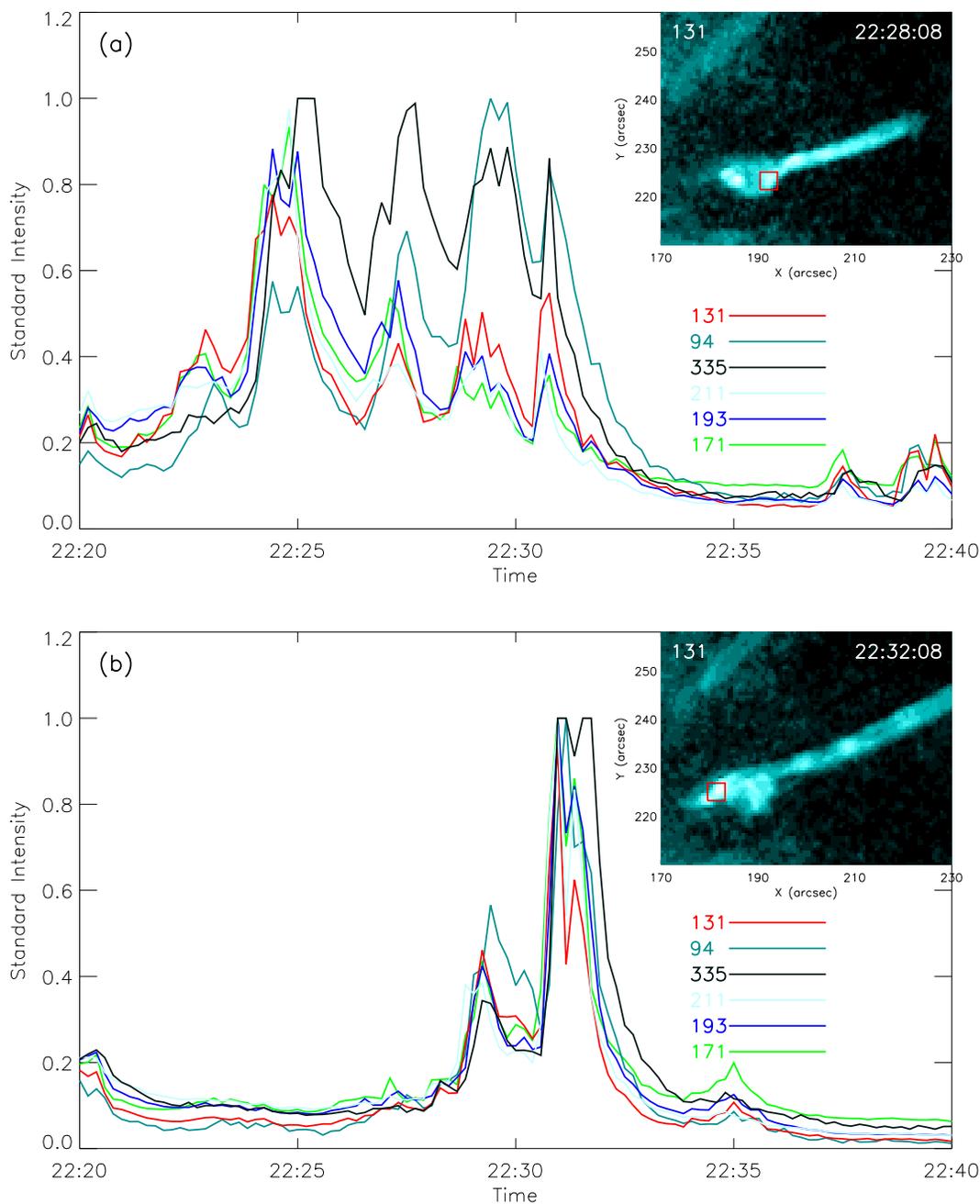}
\caption{(a) Intensity variation of the jet footpoint of Phase 1 in six different wavelengths. Top-right: An image of Phase 1 of the jet obtained from SDO/AIA 131 \AA\ at 22:28:08 UT, the red box marks the region of the footpoint in Phase 1. (b) Intensity variation of the jet footpoint of Phase 2 in six different wavelengths. Top-right: An image of Phase 2 of the jet obtained from SDO/AIA 131 \AA\ at 22:32:08 UT, the red box marks the region of the footpoint in Phase 2.\label{fig1}}
\end{figure}

\begin{figure}
\epsscale{0.90}
\plotone{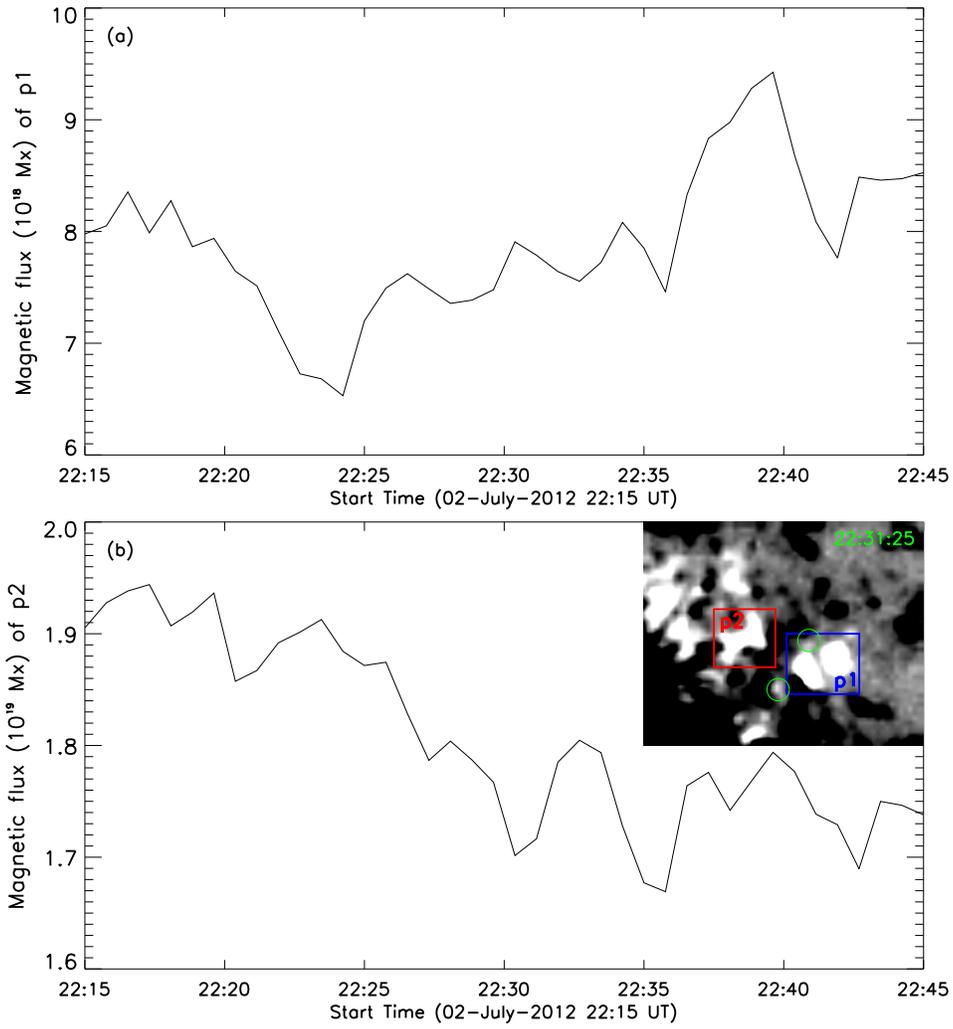}
\caption{Variation of magnetic flux in regions (a) p1 and (b) p2 from 22:15 UT to 22:45 UT on July 2nd, 2012. Top-right: HMI LOS 
magnetogram at 22:31 UT, FOV: $36^{\prime\prime} \times 28.8^{\prime\prime}$. \label{fig1}}
\end{figure}

\begin{figure}
\epsscale{1.0}
\plotone{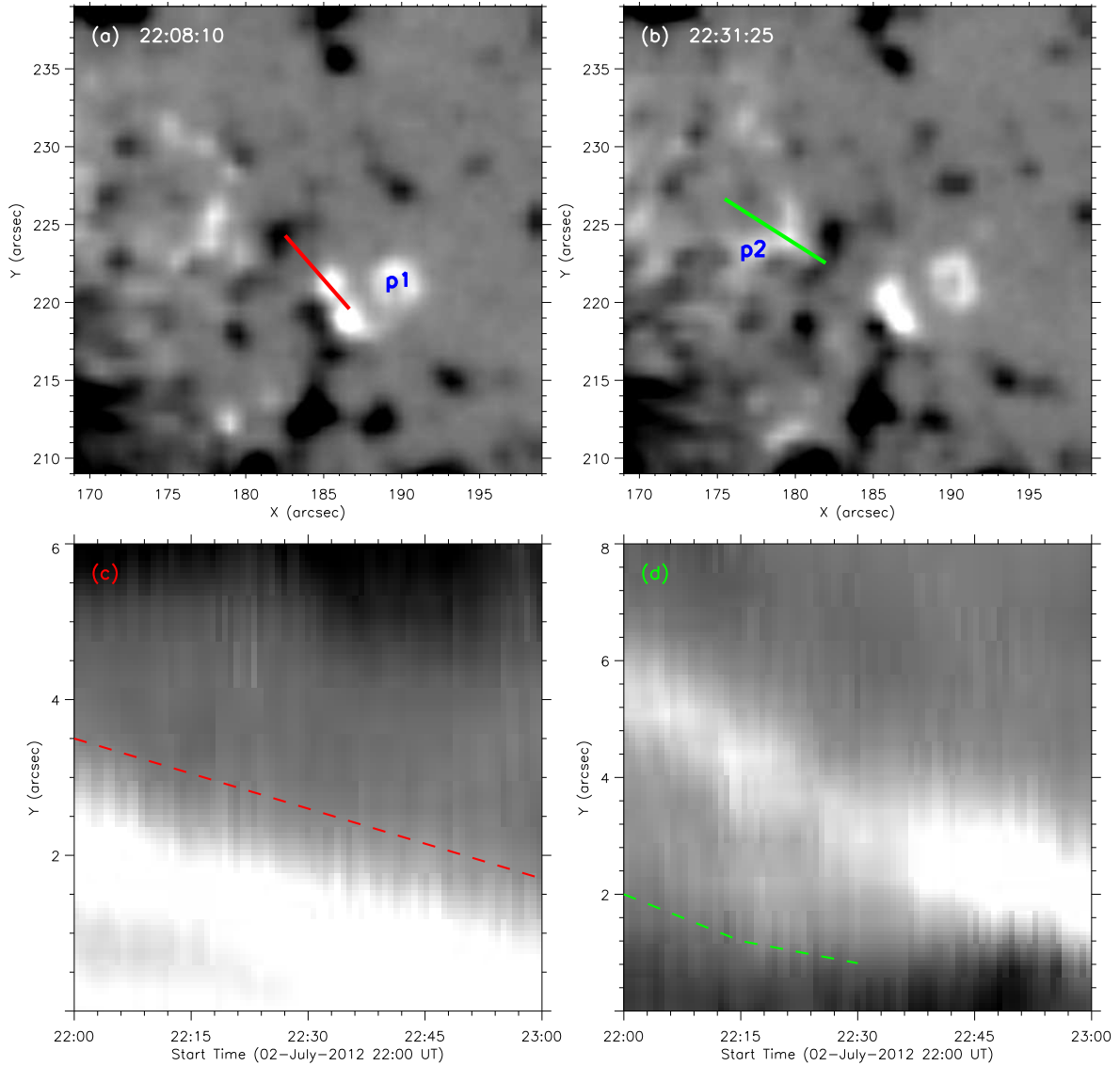}
\caption{Time slice images of regions p1 and p2 from 22:00 UT to 23:00 UT. (a) and (b) HMI LOS magnetograms at 22:08 UT and 22:31 UT, respectively. 
The dynamic range for the magnetic field strength is 200 G. (c) Time distance image for the red slit. The red dashed line shows the variation of magnetic field of p1. (d) Time distance image for the green slit.  
The green dashed line shows the variation of magnetic field of p2.\label{fig1}}
\end{figure}

\begin{figure}
\epsscale{0.8}
\plotone{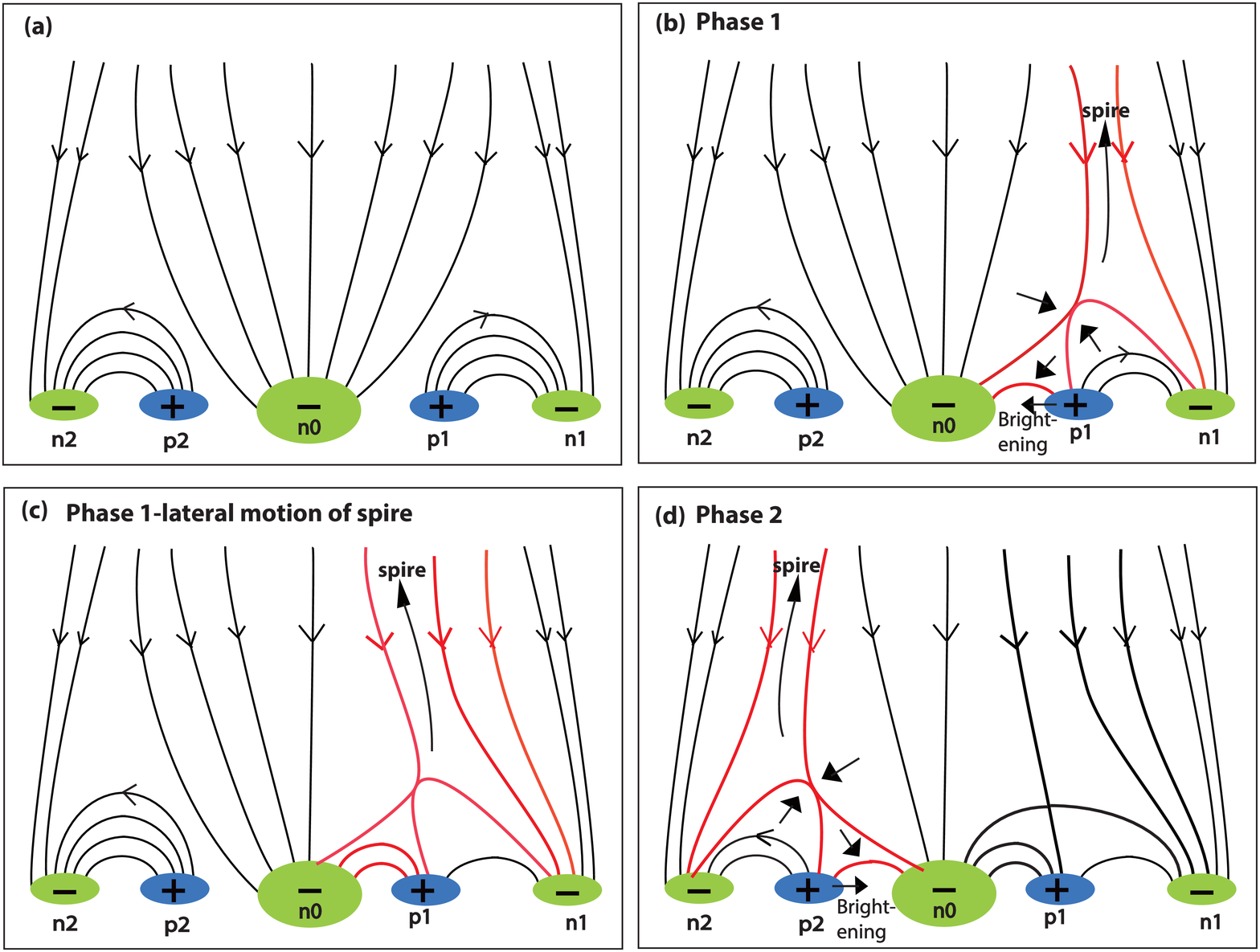}
\caption{Schematic of the topology and magnetic reconnection in the two successive phases of a complex jet showing a few representative field lines marked by 
light-headed arrows before and after bursts of reconnections. (a) Before reconnection n0 and the outer parts of n1 and n2 are open magnetic fields with negative polarity. On the both sides of n0, there are closed magnetic fields with positive polarities p1 and p2 linking to the inner parts of the negative polarities n1 and n2, as shown.  Arrows mark the direction of magnetic field lines. (b) Phase 1 of the reconnection process driven by magnetic cancellation between n0 and p1. Reconnection takes place at the 
X-point, with plasma velocities shown by solid-headed arrows. The upwards-moving outflow jet from the reconnection site is channeled upwards by the open field 
lines as a spire. The separatrices and the field lines which have reconnected (a new open field line and a new closed field line) are marked in red. The location of a chromospheric brightening is also indicated below the reconnection location and straddling n0 and p1. (c) Phase 1 continues with more reconnection at the X-point and so creates a lateral motion of the spire. (d) Phase 2 repeats the same process, but at the X-point on the left side of n0, driven by magnetic cancellation as p2 approaches n0. The separatrices and the newly reconnected field lines are shown in red.\label{fig1}}
\end{figure}

\clearpage

\end{document}